\documentclass{article}

\usepackage{arxiv}

\usepackage{cite}
\usepackage[utf8]{inputenc}
\usepackage[T1]{fontenc}
\usepackage{hyperref}
\usepackage{url}
\usepackage{amsfonts}
\usepackage{nicefrac}
\usepackage{microtype}
\usepackage{lipsum}	
\usepackage{graphicx}
\usepackage{amsmath}
\usepackage{multirow}
\usepackage{gensymb}

\title{Using neural networks to predict icephobic performance}

\author{ \href{https://orcid.org/0000-0002-2668-1938}{\includegraphics[scale=0.06]{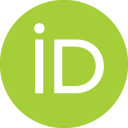}\hspace{1mm}Rahul Ramachandran}\\
	Department of Mechanical Engineering\\
	University of Nevada, Reno\\
	Reno, NV 89557 \\
	\texttt{r.ramacha6@gmail.com} \\
}

\hypersetup{
pdftitle={NN icephobicity},
pdfsubject={cs.LG, cond-mat.mtrl-sci},
pdfauthor={Rahul Ramachandran},
pdfkeywords={icephobicity, artificial neural network, mixture density network, concrete, permutation importance}
}

\begin{document}
\maketitle

\begin{abstract}
Icephobic surfaces inspired by superhydrophobic surfaces offer a passive solution to the problem of icing. However, modeling icephobicity is challenging because some material features that aid superhydrophobicity can adversely affect the icephobic performance. This study presents a new approach based on artificial neural networks to model icephobicity. Artificial neural network models were developed to predict the icephobic performance of concrete. The models were trained on experimental data to predict the surface ice adhesion strength and the coefficient of restitution (COR) of water droplet bouncing off the surface under freezing conditions. The material and coating compositions, and environmental condition were used as the models’ input variables. A multilayer perceptron was trained to predict COR with a root mean squared error of 0.08, and a 90\% confidence interval of [0.042, 0.151]. The model had a coefficient of determination of 0.92 after deployment. Since ice adhesion strength varied over a wide range of values for the samples, a mixture density network was model was developed to learn the underlying relationship in the multimodal data. Coefficient of determination for the model was 0.96. The relative importance of the input variables in icephobic performance were calculated using permutation importance. The developed models will be beneficial to optimize icephobicity of concrete.
\end{abstract}


\section{Introduction}
Ice formation on surfaces can lead to undesirable consequences such as slippery roads and walkways, downed power and communication lines, and transportation delays. Icephobic surfaces have been developed to combat the problem of icing \cite{Kim2012,Subramanyam2013, Eberle2014,Irajizad2016,Golovin2016,Kreder2016,Chen2017,Golovin2019,Hou2020}. Icephobicity of a surface has three aspects. The first is the ability to repel incoming water droplets before they can freeze on the surface. For this, the contact time of incoming water droplets with the surface needs to be minimized. The second aspect is low ice adhesion to the surface, so that any ice formed can easily be dislodged from the surface. The third aspect is to suppress the frost formation on the surface \cite{Ramachandran2016a}. Although icephobicity is closely related to superhydrophobicity, surfaces that exhibit high contact angle and small contact angle hysteresis do not always exhibit icephobicity \cite{Nosonovsky2012}. Surface textures that promote superhydrophobicity may provide sites for heterogenous ice nucleation resulting in icing \cite{Varanasi2010}. This makes the relationship between surface roughness, contact angle, and icephobicity complex to model.
\par
Artificial neural networks \cite{McCulloch1943,Rosenblatt1958,Hopfield1982} inspired by biological neural network were developed to learn underlying relationships between variables in data. Standard artificial neural networks consist of connected neurons arranged in layers or stages. Multilayer perceptrons are a type of feed-forward neural network consisting of one or more hidden layer of neurons \cite{Gardner1998}. These networks have been successfully used to build complex nonlinear predictive models for superhydrophobicity \cite{Moghadam2013,XZhang2018,Taghipour2018,JZhang2020}. However, no previous studies have used artificial neural network to model the icephobicity of surfaces. In this study, multilayer perceptrons are trained to predict the icephobic performance of biomimetic concrete surfaces.
\par
Dry concrete is porous, hydrophilic, and prone to ice accretion. Water absorbed by the concrete undergoes freeze-thaw cycle in winter weather, and forms cracks in the concrete. Also, anti-icing or deicing agents which have environmental impacts \cite{Fay2012} are regularly used on concrete surfaces in winter. Biomimetic hydrophobic and icephobic concrete was developed to tackle these issues \cite{Sobolev2013}. It was shown that material and coating composition affects the surface micro/nanotopography of concrete, and thus its icephobic performance \cite{Ramachandran2016e}. Although superhydrophobic surfaces can exhibit icephobicity, the relationship between hydrophobicity and icephobicity is not straightforward. Neural network models can be used to capture the nonlinear relationship between material composition and icephobic performance. 
\par
In this paper, we develop neural network models to predict two metrics of icephobicity of concrete\textemdash the ability to repel incoming water droplets, and the ice adhesion strength\textemdash based on material composition and environmental condition. Water droplets impinging with kinetic energy, for example, raindrops falling on a surface can either stick to the surface, bounce off, or break up into smaller droplets. In cold conditions the droplets that stick to the surface can freeze and result in ice accretion. An icephobic surface should ideally be able to repel such incoming droplets before they have a chance to accumulate on the surface and freeze. Also, the adhesion strength of ice to the surface should be low so that any ice can be easily removed by wind or vibrations. The experimental data characterizing these two aspects from our previous studies \cite{Ramachandran2016a,Ramachandran2015} are used to train, validate and test the neural network models. In the following section we discuss preprocessing the experimental data, and the topology of the neural network models.
\par

\section{Neural network models}
Two feed-forward artificial neural network models were developed and trained to predict icephobic performance of concrete surfaces. The first model predicts the effectiveness of concrete surfaces at repelling incoming water droplets measured in terms of the coefficient of restitution (COR) of droplet rebound. The second network predicts the ice adhesion strength (kPa) on concrete surfaces.
\par
\subsection{Experimental data}
Concrete samples with various mortar and coating compositions were prepared as described in \cite{Ramachandran2016a}. Mortar composition is described using four variables, namely water to cement ratio, sand to cement ratio, superplasticizer content (\% cement), and polyvinyl alcohol (PVA) fiber content (\% by volume). Water repellency is obtained by coating the samples with hydrophobic emulsions. The hydrophobic emulsion composition is described using two variables—polymethylhydroxysilane (PMHS) and silica fume content (\% by weight for both).\par
Two aspects of icephobicity of concrete were studied. First, the ability of concrete to repel incoming water droplets were studied as described in \cite{Ramachandran2015}. The COR of water droplets (approximately 3 mm in diameter) bouncing off concrete surfaces at various velocities were determined at -5 \degree C. The features and target for that dataset are listed in Table~\ref{tab:tab1}. Second, the ice adhesion strength on concrete samples were determined \cite{Ramachandran2016a} at $0 \pm 2$ \degree C. The features and target for that dataset are listed in Table~\ref{tab:tab1}.\par
\subsection{Preprocessing data}
The droplet bouncing tests generated a dataset (DS1) of 85 samples, with seven features and one target. The features are the material composition parameters and the Weber number of incoming water droplets. The target is the mean COR of the rebounding water droplets. The ice adhesion tests generated a dataset (DS2) of 30 samples, with six features. The target is the mean ice adhesion strength (kPa) on concrete surface. Datasets were scaled so that any variable with a large order of magnitude does not overwhelm the network weights during model training. Additionally, scaling improves the model performance. A holdout set of 20\% and a validation set of 20\% were randomly extracted from DS1 before feature scaling. The remaining features from DS1 were scaled. Twenty percent of data was separated from DS2 as validation set before scaling. The datasets were scaled using the Yeo-Johnson power transformer \cite{Yeo2000}, as several features were not normally distributed.\par

\begin{table}[htbp]
\begin{center}
\caption{Features and targets for the neural networks.}
\begin{tabular}{|l|l|l|} 
\hline
\multirow{2}{*}{Model: ANN} & Features & \begin{tabular}[c]{@{}l@{}}\begin{tabular}{@{\labelitemi\hspace{\dimexpr\labelsep+0.5\tabcolsep}}l}PMHS (\% wt.)\\Silica fume (\% wt.)\\Water to cement ratio\\Sand to cement ratio\\Superplasticizer (\% cement)\\PVA fibers (\% vol)Weber number\end{tabular}\end{tabular}  \\ 
\cline{2-3}
                            & Target   & \begin{tabular}{@{\labelitemi\hspace{\dimexpr\labelsep+0.5\tabcolsep}}l}Coefficient of restitution  \end{tabular}                                                                                                                                                             \\ 
\hline
\multirow{2}{*}{Model: MDN} & Features & \begin{tabular}[c]{@{}l@{}}\begin{tabular}{@{\labelitemi\hspace{\dimexpr\labelsep+0.5\tabcolsep}}l}PMHS (\% wt.)\\Silica fume (\% wt.)\\Water to cement ratio\\Sand to cement ratio\\Superplasticizer (\% cement)\\PVA fibers (\% vol)\end{tabular}\end{tabular}              \\ 
\cline{2-3}
                            & Target   & \begin{tabular}{@{\labelitemi\hspace{\dimexpr\labelsep+0.5\tabcolsep}}l}Ice adhesion strength (kPa)  \end{tabular}                                                                                                                                                            \\
\hline
\end{tabular}
\label{tab:tab1}
\end{center}
\end{table}

\subsection{Model for COR}
\begin{figure}[htbp]
\begin{center}
\includegraphics[width=6.5in]{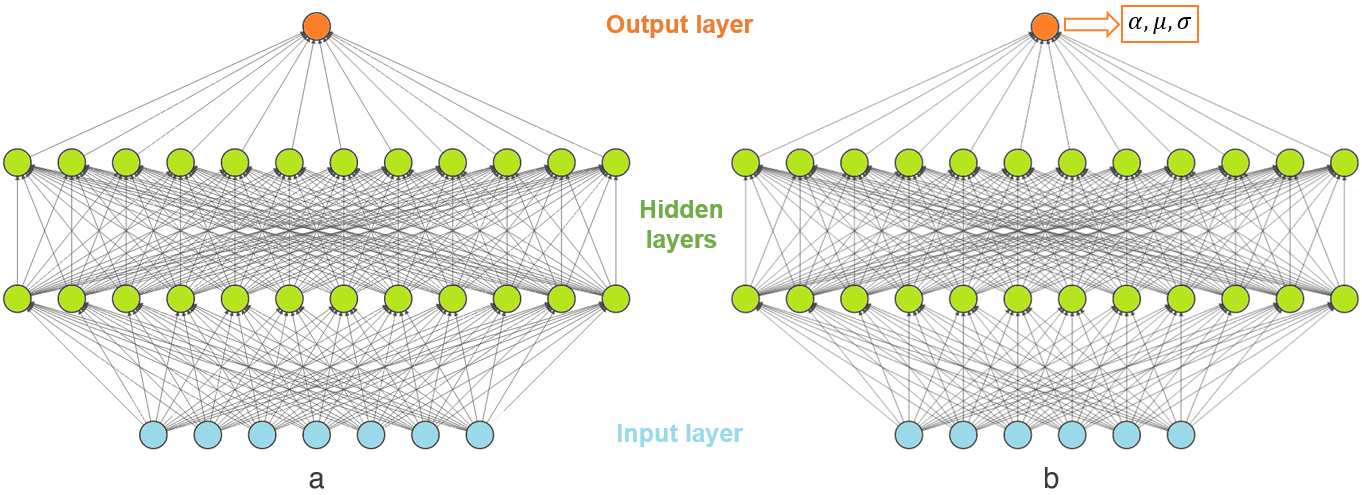}
\caption{Neural network topologies. (a) ANN for predicting COR (b) MDN for predicting ice adhesion strength.}
\label{fig:fig1}
\end{center}
\end{figure}

The topology of the artificial neural network (ANN) for predicting COR is shown in Figure\ref{fig:fig1}a. The network has seven input neurons, two hidden layers with 12 neurons each, and one output neuron. Activation functions are used to introduce nonlinearity in the model. The Rectified Linear unit (ReLU) f(x) = max(0,x) is used as the activation function for the hidden layer. ReLU offers several computational advantages over other activation functions \cite{LeCun2015,Dahl2013,Nair2010,Nwankpa2018}. For the output layer, a linear activation function is used.

The model was evaluated using the bootstrap method \cite{Tibshirani1996}, with root mean squared error (RMSE) as the metric, and adaptive moment estimation optimization \cite{Kingma2014}. RMSE is given by eq.~\ref{eq:eq1} where N is the number of samples, ŷ and y are the predicted and target values, respectively. The mean value of the error metric, its confidence interval, and the mean number of epochs to train the model were determined by iteratively resampling the dataset 500 times with replacement. The final network model was trained and validated over the mean number of epochs estimated by the bootstrap method. The holdout (test) set was used to estimate the model performance after deployment.

\begin{equation}
RMSE =\sqrt{\frac{1}{N}\sum_{i=1}^{N}(\hat{y}_{i} - y_{i})^2}
\label{eq:eq1}
\end{equation}

\subsection{Model for ice adhesion strength}
\begin{figure}[htbp]
\begin{center}
\includegraphics[width=3.25in]{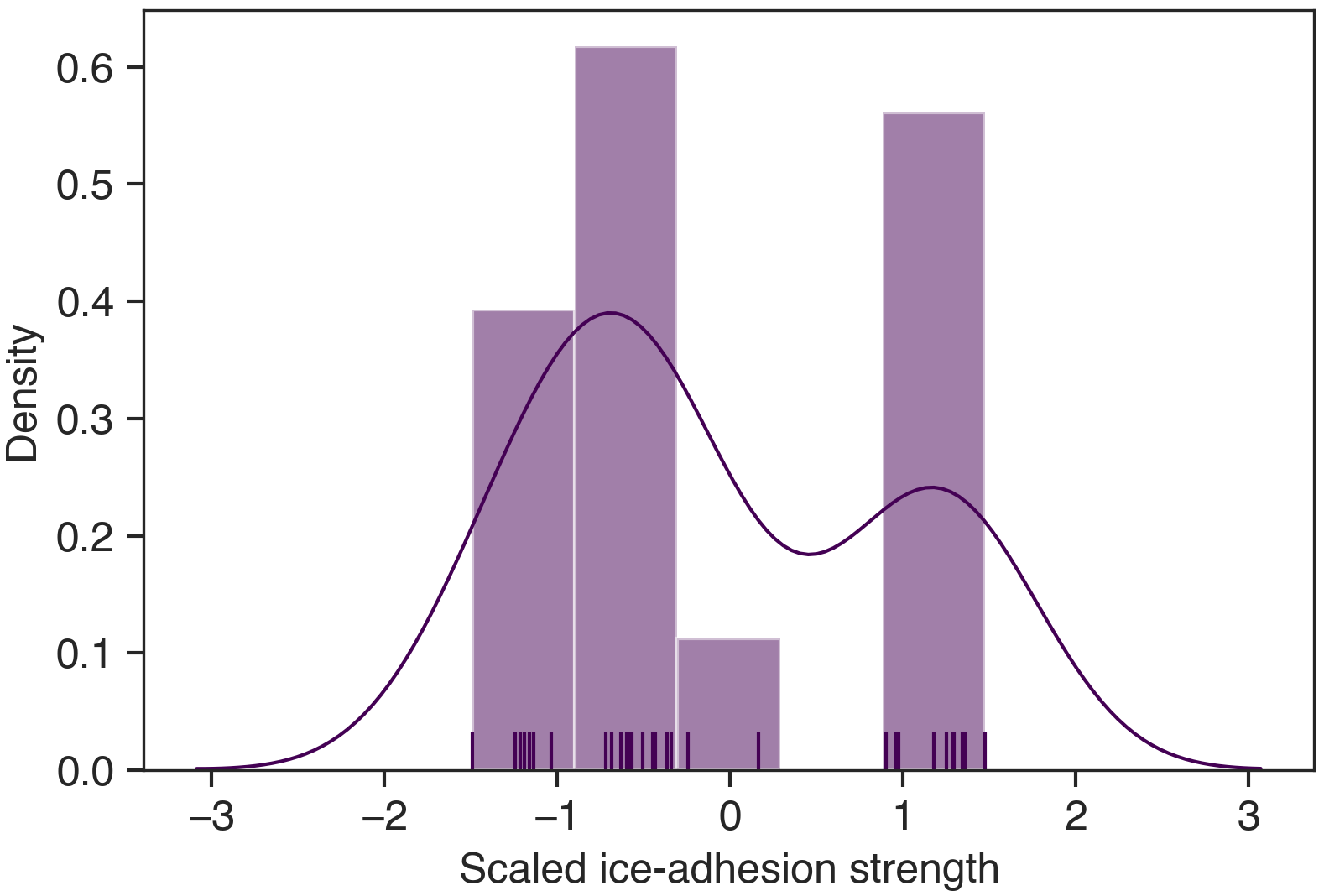}
\caption{Ice adhesion strength values exhibit a bimodal distribution.}
\label{fig:fig2}
\end{center}
\end{figure}

DS2 consists of a range of experimental data using both hydrophobic and regular concrete. Because of the wide range of ice adhesion values, the target in DS2 exhibits a bimodal distribution (Figure \ref{fig:fig2}). Hence a mixture density network (MDN) was used to model the ice adhesion of concrete. MDN is a combination of a feed-forward neural network and a Gaussian mixture model (GMM) \cite{Bishop1994}. Conditional probability density of the target (\textit{t}) in terms of the feature (\textit{x}) is a GMM of the form

\begin{equation}
p(t|x) =\sum_{i=1}^{m}\alpha_i(x)N(\mu_i(x),\sigma_i^2(x);t), \sum_{i=1}^{m}\alpha_i(x)=1, \alpha_i \ge 0
\label{eq:eq2}
\end{equation}

where m is the number of mixture components, $\alpha_{i}$ is the mixing coefficient, and N denotes a normal distribution with mean $\mu_{i}$  and variance $\sigma_{i}^2$. The feed-forward network can be used to estimate the mixture parameters.\par
Figure \ref{fig:fig1}b shows the topology of MDN. The network has six input neurons and two hidden layers with 12 neurons each. The output of MDN is a parameter vector consisting of mixing coefficients, means, and standard deviations. ReLU activation functions were used for the two hidden layers. Since the mixing coefficients for MDN are positive numbers which add up to one, a softmax \cite{Bridle1990} activation function was used. The MDN learning is driven by the minimizing the loss function,

\begin{equation}
L =\sum_{q}-log\{\sum_{i=1}^{m}\alpha_i(x)N(\mu_i(x),\sigma_i^2(x);t\}
\label{eq:eq3}
\end{equation}

where q is the total number of patters in the training set \cite{Bishop1994}. The predictive models were generated using the opensource machine learning packages TensorFlow \cite{Abadi2015} and scikit learn \cite{Pedregosa2011}. The evaluation results and the predictive performance of the models are discussed in the following sections.

\subsection{Variable importance}
Relative importance of the predictors in the models were determined using permutation importance \cite{Breiman2001}. Values of an input variable were randomly perturbed, while all the remaining variables were left intact. The model was then used to predict on the perturbed dataset and the results were compared to the baseline results from the trained model on the unperturbed dataset. The decrease in model prediction quality was monitored. This process was repeated several times for each variable to obtain a mean value of degradation in prediction quality.
\section{Results}
\begin{figure}[htbp]
\begin{center}
\includegraphics[width=6.5in]{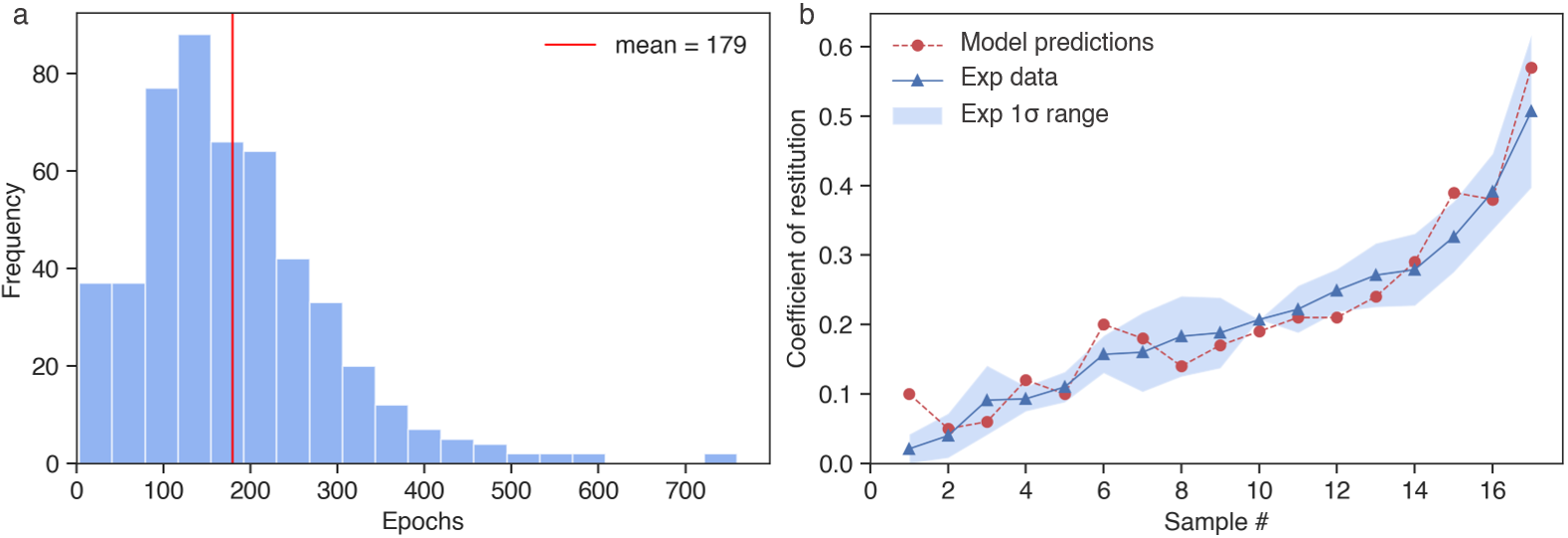}
\caption{(a) Histogram showing the results of bootstrap method. The mean number of epochs to train the model was 179. (b) Comparison of neural network COR predictions with experimental (test) data. Majority of the predictions are within one standard deviation of the experimental values.}
\label{fig:fig3}
\end{center}
\end{figure}

\begin{figure}[htbp]
\begin{center}
\includegraphics[width=6.5in]{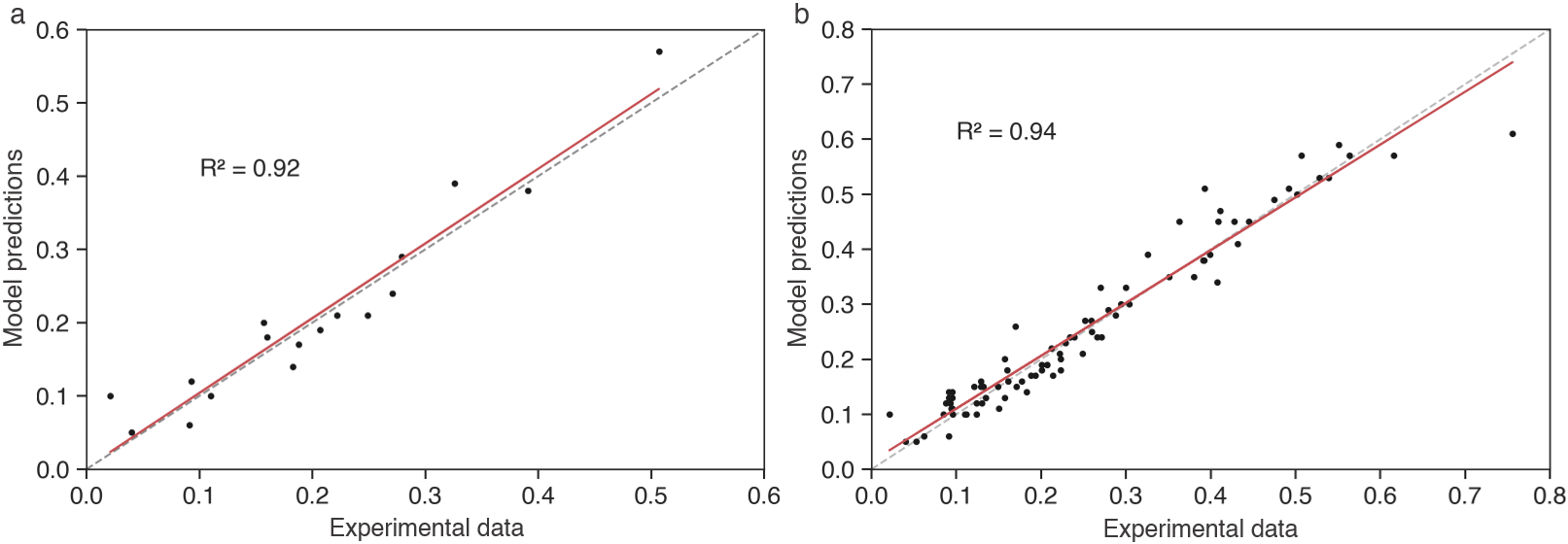}
\caption{(a) Goodness-of-fit of COR predictions with the holdout set. (b) Comparison of COR predictions with entire dataset of experimental values.}
\label{fig:fig4}
\end{center}
\end{figure}

After training and validating ANN 500 times using bootstrap method, its generalization performance was obtained. The mean number of epochs for the network was obtained as 179 (Figure \ref{fig:fig3}a). The 90\% confidence interval for RMSE was [0.042, 0.151], with a mean of 0.08. The final network was trained for 179 epochs, and the predictions for the experimental data is shown in Figure \ref{fig:fig3}b. It was observed that majority of the predictions were within one standard deviation of the experimental data. The coefficient of determination R\textsuperscript{2} for the test set was 0.92 (Figure \ref{fig:fig4}a). The model predictions on the entire experimental dataset is shown in Figure \ref{fig:fig4}b.\par

\begin{figure}[htbp]
\begin{center}
\includegraphics[width=6.5in]{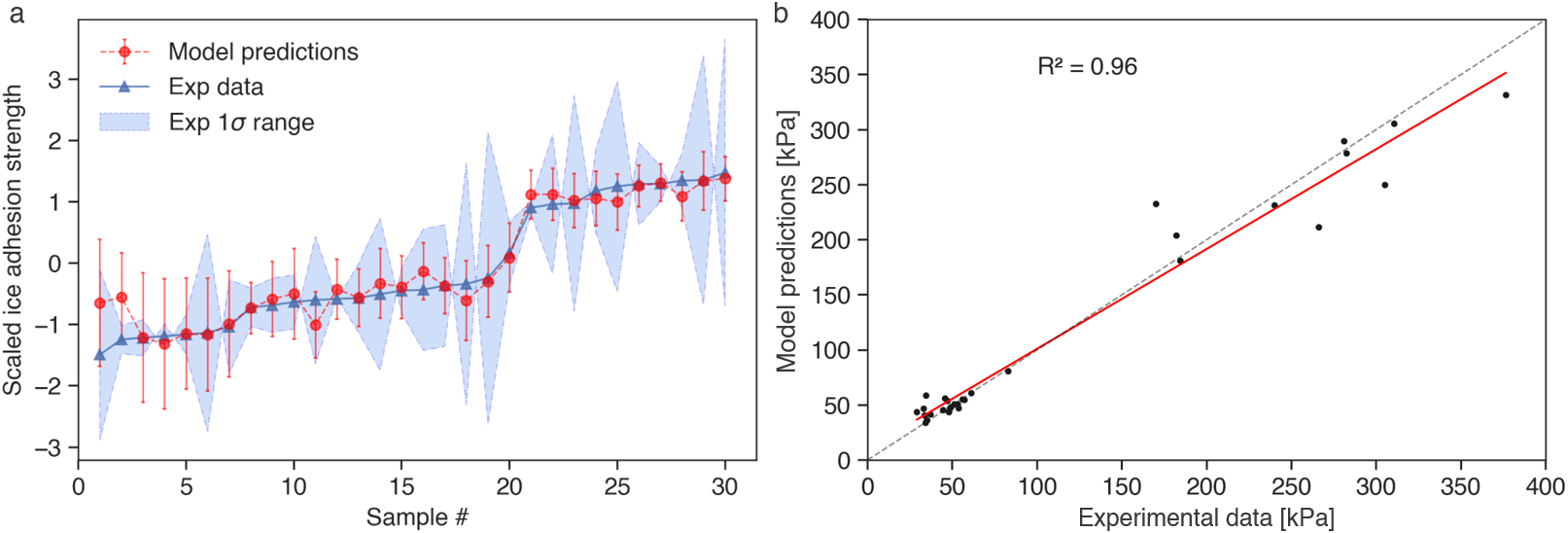}
\caption{(a) Scaled results of MDN model compared with scaled experimental data. (b) Comparison of ice adhesion strength predictions to experimental data.}
\label{fig:fig5}
\end{center}
\end{figure}

The MDN predicts mixing coefficients, means and standard deviations of two Gaussian mixtures. Plot in Figure \ref{fig:fig5}a shows the scaled ice adhesion strength on concrete samples. The scaled experimental data and corresponding one standard deviation range is denoted in blue. The predictions and corresponding standard deviations are denoted with red dotted line. Majority of the predictions are within the one standard deviation range of the experimental data. Figure \ref{fig:fig5}b shows the comparison between the model predictions and the entire experimental dataset. R\textsuperscript{2} for the entire dataset was 0.96. These results demonstrate that the model predicted well on the dataset.\par

\begin{figure}[htbp]
\begin{center}
\includegraphics[width=6.5in]{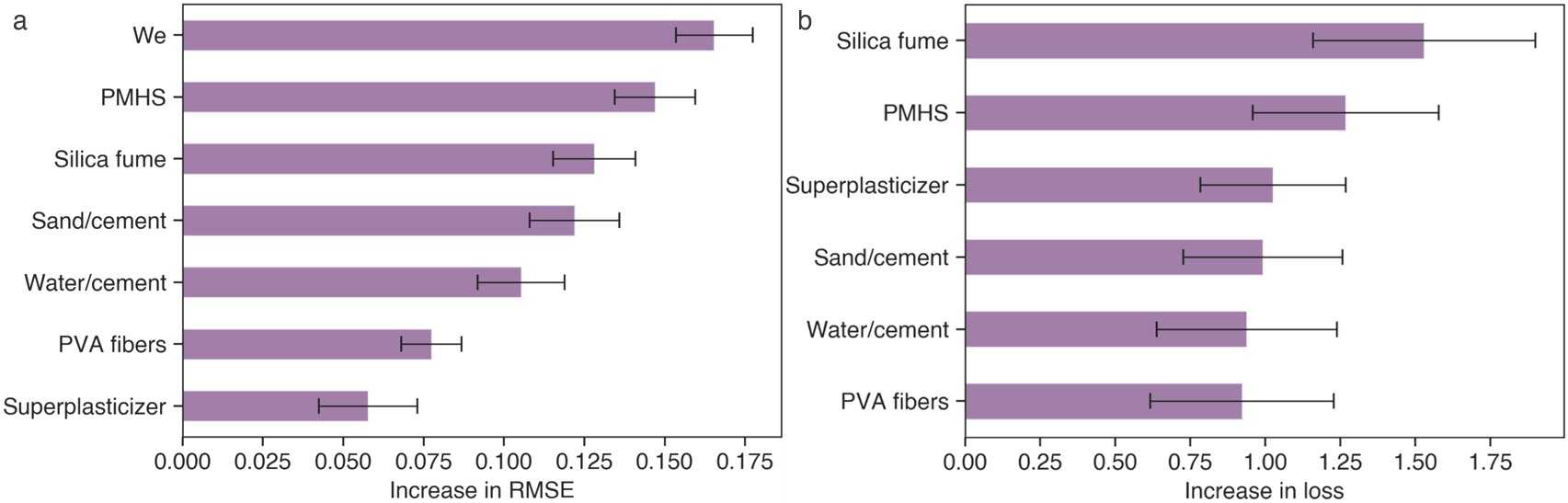}
\caption{Permutation importance of features. (a) Predictors of ANN ranked by the increase in root mean squared error after perturbing. (b) Predictors of MDN ranked according to the increase in the loss function after perturbing.}
\label{fig:fig6}
\end{center}
\end{figure}

The relative importance of the features in the two models are as shown in Figure \ref{fig:fig6}.  Each variable was perturbed 20 times to obtain the mean and standard deviation of the model’s error metric. Predicting using a trained model on a perturbed dataset will cause an increase in the error from the baseline value. Higher the error, more important is the feature to the model. Increase in RMSE was considered for ANN (Figure \ref{fig:fig6}a). Weber number was found to be the most important feature, followed by the coating composition (PMHS and silica fume content), and the concrete composition. For MDN, the loss function (eq.~\ref{eq:eq3}) was used to compare the degradation in model performance. The coating composition was once again the most important feature for the model (Figure \ref{fig:fig6}b).

\section{Discussion}
The models developed in this study are used to predict two aspects of icephobicity\textemdash the ability to repel incoming water droplets, and the ice adhesion strength\textemdash of concrete. The first aspect is characterized by the COR of water droplets impacting the surface. Surfaces with high contact angle can repel incoming water droplets. However, in some cases the droplets can get pinned to hydrophobic/superhydrophobic surfaces. The effectiveness of the surfaces in repelling incoming water droplets depends also on the surface micro/nanotopography \cite{Ramachandran2015,Bird2013}. Constituents of the cementitious composite, such as sand and PVA fibers provide micro/nanotopography to the concrete surface. Hence, their compositions are important features to be considered for the neural network. Thus, modeling COR metric is not straightforward. This is where the versatile statistical method of multilayer perceptron becomes useful. \par
The training for ANN was performed with 51 datapoints. In training phase, the weights and biases are repeatedly updated by backpropagation. During training the model can pick up the noise in the data instead of learning the underlying relationship between the features and target. To avoid this, a validation set is used to monitor the error metric. If during training, the validation error increases while the train error stays the same or continues to decrease, then the model is overfitting. To prevent overfitting, the error metrics were monitored for the 179 epochs. Early stopping can also be used to prevent overfitting. \par
Data leakage from validation and test sets to training set can result in overly optimistic predictive models. Data leakage was minimized by separating the train, test, and validation data before scaling. The holdout test set of size 17 was used to check if the final model predictions were overly optimistic. R\textsuperscript{2} of 0.92 for the test set compared to 0.94 for the entire dataset implies that the model was able to learn the underlying relationship in the data over the entire range of COR. Note that the error produced in the model can be reduced by changing the hyperparameters, but at the cost of overfitting due to limited number of samples. \par
Permutation importance (Figure \ref{fig:fig6}a) estimated that Weber number was the most important variable in the model. Note that Weber number is an environmental variable and unrelated to the material composition. The most important variable related to the material itself was PMHS content in the coating. Silica fume content in the coating is the next significant variable, followed by the sand to cement ratio. The importance of these can be explained as follows. Hydrophobic agent PMHS increases the water contact angle on the concrete surface. Silica fume and sand particles provide micro-scale roughness, thus enhancing the hydrophobicity of the surface. Asymmetric surface roughness can reduce pinning of incoming water droplets on the surface \cite{Ramachandran2015,Bird2013}. Pinning implies an inelastic collision of droplet with the surface, and thus low COR. \par
The second aspect of icephobicity modeled is the ice adhesion strength on concrete surface. The adhesion strength depends not only on the water-repellency but also the surface micro/nanotopography. As discussed previously, material and coating parameters need to be considered as features for the model. Since the experimental dataset consisted of 30 samples, it was split for train and validation. A holdout set was not available for this model. The validation set and early stopping strategy were used to prevent overfitting of the model. In early stopping the training is stopped when the model stops improving on the validation set. Data leakage was minimized by scaling the train data separately.\par
MDN was trained on a bimodal dataset, and thus predicts ice adhesion strength of both icephobic and regular concrete. Icephobic concrete exhibits lower ice adhesion strength compared to regular concrete. From Figure \ref{fig:fig5}b it can be observed that the MDN model was relatively better at predicting lower values of ice adhesion strength. This can be attributed to the fact that 2/3 of the data are of icephobic concrete samples. A randomly picked training set is likely to contain more experimental values of icephobic concrete. This can be rectified by adding more experimental data to the dataset. Note that MDN provided better results than a regular feed-forward neural network. The versatility of MDN allows compatibility with a multimodal dataset. \par
Silica fume content was the most significant variable as estimated using permutation importance for the MDN model. Note that silica fume provides a smaller scale roughness compared to the sand particles. PMHS content in the coating was the next significant variable because of its functionality. PMHS provides high water contact angle, thus minimizing the area of contact between water and the surface. The roughness provided by silica fume further enhances the contact angle by creating pockets of trapped air at the interface. The air pockets and low contact area result in weak adhesion of ice to the surface. \par
The models presented in the paper predict the metrics for icephobic performance of engineered surfaces. These models can be extended for other targets such as material properties by using additional predictors in the training and validation phases. MDN is ideal for modeling materials where multimodal data can be expected, such as hydrophilic/hydrophobic/superhydrophobic materials.
\par
\section{Conclusion}

Two artificial neural network models were developed to predict the icephobic performance of concrete. The first model predicts the ability of concrete to repel incoming water droplets at  -5 \degree C using COR of water droplet rebound as the metric. The first model has RMSE of 0.08, with a 90\% confidence interval of [0.042, 0.151]. R\textsuperscript{2} for holdout data was 0.92, demonstrating the model’s robust predictive performance over the entire range of dataset. The model was able to capture the effect of material and coating compositions, and environmental conditions on the collision dynamics of water droplets with the concrete surface. The second model is a mixture density network to predict the adhesion strength of ice on concrete at 0 \degree C. Early stopping was used to prevent model overfitting. The model captured the effect of material and coating compositions to predict the icephobicity of both icephobic and regular concrete samples. R\textsuperscript{2} for the entire dataset was 0.96. Variables associated with coating composition were found to be relatively more important than the ones associated with concrete composition in both models. The versatility of MDN makes it suitable for material modeling where multimodal data can be expected.\par

\end{document}